\documentclass[aps,pre,preprint,superscriptaddress,showpacs]{revtex4-1}
\usepackage[utf8]{inputenc}
\usepackage[english]{babel}
\usepackage{amsmath}
\usepackage{amsfonts}
\usepackage{amssymb}
\usepackage{amsthm}
\usepackage{graphicx}
\usepackage{enumerate}
\usepackage{color}
\usepackage{booktabs}
\usepackage{bbm}
\usepackage{mathrsfs}
\usepackage{graphicx}
\usepackage{bm}

\newcommand{\eps}{\varepsilon}

\begin{document}

\title{Steady state and mean recurrence time for random walks on stochastic temporal networks}


\author{Leo Speidel}
\affiliation{Department of Mathematical Informatics, The University of Tokyo, Tokyo, Japan}
\affiliation{JST, ERATO, Kawarabayashi Large Graph Project, Tokyo, Japan}

\author{Renaud Lambiotte}
\affiliation{Department of Mathematics/Naxys, University of Namur, Namur, Belgium}

\author{Kazuyuki Aihara}
\affiliation{Department of Mathematical Informatics, The University of Tokyo, Tokyo, Japan}
\affiliation{Institute of Industrial Science, The University of Tokyo, Tokyo, Japan}

\author{Naoki Masuda}
\email{naoki.masuda@bristol.ac.uk}
\affiliation{Department of Engineering Mathematics, University of Bristol, Bristol, UK}
\affiliation{CREST, JST, Saitama, Japan}

\date{\today}

\begin{abstract}
Random walks are basic diffusion processes on networks and have applications in, for example, searching, navigation, ranking, and community detection. Recent recognition of the importance of temporal aspects on networks spurred studies of random walks on temporal networks. Here we theoretically study two types of event-driven random walks on a stochastic temporal network model that produces arbitrary distributions of interevent times. In the so-called active random walk, the interevent time is reinitialized on all links upon each movement of the walker. In the so-called passive random walk, the interevent time is reinitialized only on the link that has been used the last time, and it is a type of correlated random walk. We find that the steady state is always the uniform density for the passive random walk. In contrast, for the active random walk, it increases or decreases with the node's degree depending on the distribution of interevent times. The mean recurrence time of a node is inversely proportional to the degree for both active and passive random walks. Furthermore, the mean recurrence time does or does not depend on the distribution of interevent times for the active and passive random walks, respectively.
\end{abstract}

\pacs{64.60.aq, 05.40.Fb, 89.75.Hc}

\maketitle

\section{Introduction}

A broad range of diffusive processes on networks, from consensus formation \cite{Liggett1985book,Durrett1988book} to current flows on electric circuits~\cite{Doyle1984book}, can be modeled by random walks or, equivalently, by Markov chains. Their unbiased exploration of the underlying structure also makes them popular tools for designing algorithms for, e.g., navigation and search on networks \cite{Kleinberg2000Nature,Adamic2001PhysRevE,Guimera2002PhysRevLett}, defining central nodes in a given network \cite{Brin1998conf,Noh2004PhysRevLett,Newman2005SocNetw}, community detection \cite{Rosvall2008PNAS}, and respondent-driven sampling~\cite{Salganik2004SocMet,Volz2008JOffStat}. In tandem with these applications,
the impact of network structure on dynamics of random walks, including the hitting time, mixing time, and stationary density has been extensively studied.

However, recent studies identified limitations of the classical network paradigm, where dynamics is modeled by a dynamical process on a static underlying structure. In a broad range of empirical systems, evidence suggests instead that dynamics presents non-trivial correlations between events and long-tailed interevent time distributions \cite{Eckmann2004PNAS,Barabasi2005Nature,VazquezA2006PRE_burst}. These observations are incompatible with the Poissonian statistics implicitly assumed in stochastic models, therefore calling for richer models for temporal
networks \cite{HolmeSaramaki2012PhysRep}. 

An important practical question concerns the impact of the temporality of a network on diffusion. In order to address this question, a first approach consists in simulating random walks on real or synthesized data of time-stamped event sequences, and to compare their dynamical properties to those of properly defined null models \cite{Starnini2012PRE,Ribeiro2013SciRep,Scholtes2013,RochaMasuda2014NewJPhys}.
A second approach consists in studying analytically the properties of random walks on specific models of temporal networks. 
In most studies, however, network structure changes at regular time intervals, and transitions between networks at different times are independent \cite{Perra2012PhysRevLett,Ribeiro2013SciRep,Masuda2013PRL} or Markovian \cite{Avin2008LNCS,Figueiredo2012SIGMETRICS}. 

In the present work, we follow the latter path to 
analytically model diffusion under the non-Poissonian nature of interevent times. Previous studies proposed modeling temporal networks as stochastic sequences of events obeying a prescribed distribution of interevent times attached on each link~\cite{Hoffmann2013chapter,Hoffmann2012PhysRevE,Rocha2013PlosComputBiol,Delvenne2013arxiv}. A random walk process, called the active random walk, was then defined as a renewal process; i.e., after a walker arrives at a node,
the interevent times attached to all links incident to the node are reinitialized \cite{Hoffmann2013chapter,Hoffmann2012PhysRevE,Delvenne2013arxiv}. Despite its non-Markovianity owing to the fact that
the rate at which an event takes place depends on the time of the previous event, this stochastic process can be described by a generalized master equation, and some of its properties, such as the stationary density \cite{Hoffmann2013chapter} and the relaxation time \cite{Delvenne2013arxiv}, were analytically solved.
In this work, we will first derive an analytical expression for the mean recurrence time for the active random walk. Then, we consider a different non-renewal process, called the passive random walk, in which interevent times are reset only on the links traversed at each jump. 
We will use the fact that the passive random walk shows
a stronger non-Markovianity than the active random walk does because passive random walkers remember their past trajectories to some extent.
Finally, we will perform numerical simulations to test our analytical predictions and compare the stationary density and mean recurrence time between the two types of walks.

\section{Model}\label{sec:model}

\begin{figure}
\centering
\includegraphics[width=12cm]{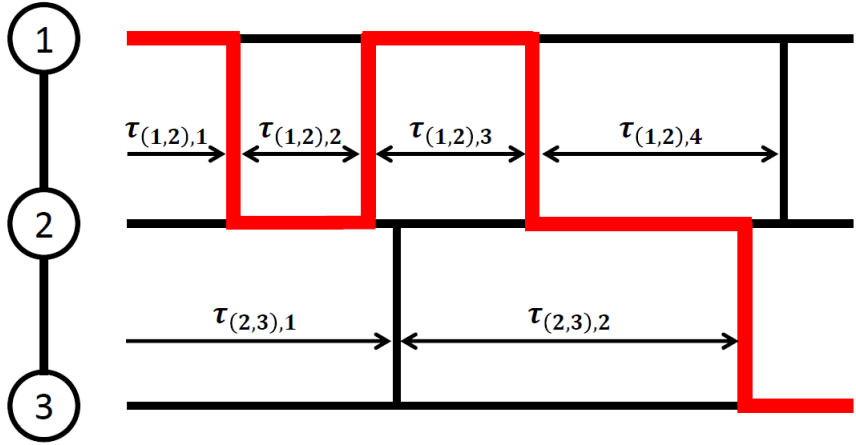}
\caption{(Color online) Schematic of the passive random walk on an example network possessing $N=3$ nodes. The walker starts at node 1 at $t=0$. The link activation is represented by horizontal bars. $\tau_{e,i}$ ($i\ge 1$) represents the $i$th interevent time on link $e$. The random walker follows the path indicated by the thick lines.}
\label{fig:schematic passive}
\end{figure}

We consider an undirected network on $N$ nodes. We denote the set of nodes by $V=\{1, 2, \ldots, N\}$ and the set of links by $E$. We often refer to this network as the aggregate network because it is considered as the aggregation of the temporal network, which we introduce in the following, across time. Stochastic temporal networks add a time dimension to the aggregated networks by assigning a random interevent time $\tau_e$ to each link $e \in E$ in a renewal manner \cite{Hoffmann2013chapter,Hoffmann2012PhysRevE}. For a chain network of three nodes, a hypothetical sequence of interevent times is depicted in Fig.~\ref{fig:schematic passive}. The interevent time, denoted by $\tau_{e,1}$, $\tau_{e,2}$, $\ldots$, for link $e$ in Fig.~\ref{fig:schematic passive}, is the interval between two consecutive activation events of the link.
We denote the probability density function (PDF) of $\tau_e$ by $\psi_e(t)$. We assume that the mean of $\tau_e$ is finite.

Link activation is an instantaneous event, and we assume that the random walker moves from the current node to the neighboring node through the activated link whenever it is possible. The example shown in Fig.~\ref{fig:schematic passive} indicates that, if the walker starts from node 1 at $t=0$, it moves to node 2 at $t=\tau_{(1,2),1}$. Then, it moves back to node 1 at $t=\tau_{(1,2),1}+\tau_{(1,2),2}$.  After coming back to node 2 at $t=\tau_{(1,2),1}+\tau_{(1,2),2}+\tau_{(1,2),3}$, the walker transits to node 3 at $t=\tau_{(2,3),1}+\tau_{(2,3),2}$. Some remarks are in order. First, some activation events, such as those immediately following the interevent times $\tau_{(1,2),4}$ and $\tau_{(2,3),1}$, may not be used by a walker. Second, although the walker's movement as illustrated in Fig.~\ref{fig:schematic passive} may look deterministic, it is stochastic because the interevent times are random variables. Third, the probability that more than one links are simultaneously activated is equal to zero because the model is defined in continuous time. Fourth, the walker is assumed not to have any internal waiting time. Therefore, the walker instantaneously moves upon the activation of a link incident to the walker's location. Fifth, the continuous-time nature of our model makes it different from the previously proposed discrete-time random walk defined on the so-called activity driven model \cite{Perra2012PhysRevLett}. A main difference between the present model and those considered in Ref.~\cite{Starnini2012PRE} is that the former generates interevent times from a given PDF, which lends itself to an analytical study of  
the process, whereas the latter uses interevent times observed in real  
data (and their randomizations), more appropriate for the study of  
empirical data.

\begin{figure}
\centering
\includegraphics[width=12cm]{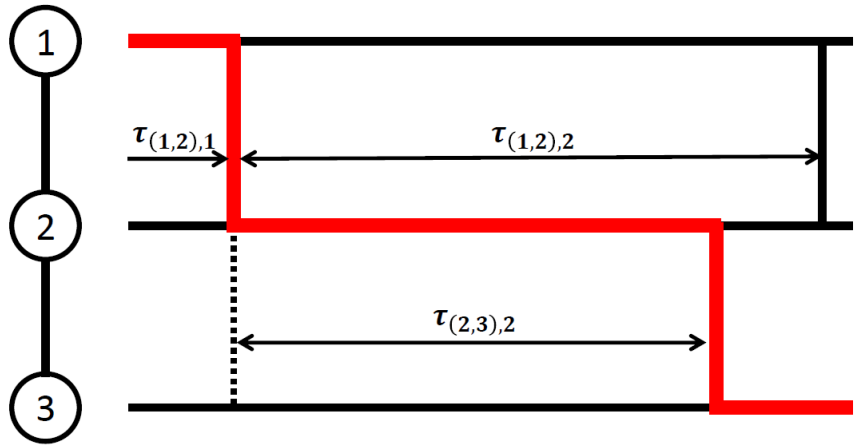}
\caption{(Color online) Schematic of the active random walk on an example network. The walker starts at node 1 at $t=0$ and follows the path indicated by the thick lines. Note that $\tau_{(2,3),2}$ is not the second interevent time on link $(2, 3)$. It is a realization of the interevent time drawn at $t=\tau_{(1,2),1}$.}
\label{fig:schematic active}
\end{figure}

We analyze two versions of random walks on stochastic temporal networks \cite{Hoffmann2012PhysRevE}.
The first version is the so-called active random walk.
In the active random walk, when the random walker arrives at a node, it reinitializes the interevent times on all links, which makes the process renewal. In this case,
the waiting time, i.e., the time for which a walker waits on a node before the link appears, is equivalent to the interevent time
\cite{Allen1990,lambiotte2013}. In the example shown in Fig.~\ref{fig:schematic active}, once the random walker moves from node 1 to node 2 at $t=\tau_{(1,2),1}$, the new interevent times, denoted by $\tau_{(1,2),2}$ and $\tau_{(2,3),2}$, are independently drawn. Because $\tau_{(1,2),2} > \tau_{(2,3),2}$, the walker moves to node 3 at $t=\tau_{(1,2),1} + \tau_{(2,3),2}$. It should be noted that interevent times for link $(2, 3)$ before $t=\tau_{(1,2),1}$ do not matter for the movement of the walker in question, reflecting the renewal nature of the active random walk.
Phenomena in which a node starts something in response to an external input or the arrival to the node might be represented by the active walk. Possible examples include the broadcasting of gossips \cite{Hoffmann2013chapter} and random surfing on the World Wide Web by users.
In the latter case, a user that has arrived at a webpage may stay there looking at the content.
Then, after a randomly distributed amount of time, the user may decide to jump to a different webpage, which is selected with equal probability among the hyperlinks that the original page has.

The second version is the so-called passive random walk, which does not assume
the reinitialization at all links. When the random walker moves to a neighbor through link $e$, a new interevent time is drawn only for $e$.
In fact, the example shown in Fig.~\ref{fig:schematic passive} corresponds to the passive random walk. Transmission of infectious diseases or information may be better described with the passive than active random walk because infection of a node does not seem to instantaneously affect interevent times on other links \cite{Hoffmann2013chapter}.
In other words, to a first-order approximation, the inter-contact time on any link $e$ can be regarded to obey a given distribution and is not altered by whether an infection occurs upon the contact. Then, a pathogen or piece of information starting from an initial node may travel to other nodes via the passive random walk.
The complexity of the passive random walk lies in its non-renewal nature. In other words, transition rates of the passive random walk depend on the trajectory that the walker has taken such that we have to account for the entire trajectory of the random walker to accurately evaluate its behavior. It should be noted
that for exponentially distributed interevent times the active and passive random walks are identical and reduce to the usual continuous-time random walk on the aggregate (i.e., static) network.

\section{Active random walk}\label{sec:active}

The steady state of the active random walk was derived through a master equation approach in Ref.~\cite{Hoffmann2012PhysRevE}. In this section, we first review these results (Secs.~\ref{sub:formulate active} and \ref{sub:steady state active}). Then, we derive the mean recurrence time for the active random walk (Sec.~\ref{sub:mean recurrence time active}). The main results when the interevent times of different links are distributed according to a common PDF are shown in Eqs.~\eqref{eq:p_i^* active} and \eqref{eq:T_{i|i} active identical} for the steady state and mean recurrence time, respectively. 

\subsection{Probability flows}\label{sub:formulate active}

We denote the probability that the random walker is located at node $i$ ($1\le i\le N$) at time $t$ by $p_i(t)$. The normalization is given by $\sum_{i=1}^N p_i(t)=1$. The rate at which the walker arrives at node $j$ from node $i$ at time $t$ is denoted by $q_{j \gets i}(t)$. The transition rate for a single walker to move from $i$ to $j$ at time $t$ is given by $r_{j \gets i}(t) := q_{j \gets i}(t)/p_i(t)$. 
The master equation that governs the random walk is given by
\begin{align}
\label{eq:master}
\frac{d}{dt}p_i(t) &= \sum_{j;(ji) \in E} \left[ q_{i \gets j}(t) - q_{j \gets i}(t) \right]\\\
&=\sum_{j;(ji) \in E} \left[ r_{i \gets j}(t)p_j(t) - r_{j \gets i}(t)p_i(t) \right]. 
\notag
\end{align}
If the underlying static network is connected, which we assume in the following, the random walk is mixing and the stationary density, denoted by
$p_i^*:= \lim_{t \to \infty} p_i(t)$, 
is obtained if we set $\lim_{t \to \infty} dp_i/dt = 0$ for all nodes $i$.

For exponentially distributed interevent times, the transition rate is 
given by
\begin{equation}
r_{j \gets i}(t) = r_{i \gets j}(t) =
\begin{cases}
\frac{1}{\langle \tau_{(ij)} \rangle} & \textrm{if } (ij) \in E,\\  0 & \textrm{if } (ij) \not\in E,
\end{cases}
\end{equation}
where $\langle \tau_{(ij)} \rangle$ denotes the mean of $\tau_{(ij)}$, the interevent time on link $(ij)$. By combining Eq.~\eqref{eq:master} and
$r_{j \gets i}(t) = r_{i \gets j}(t)$, we conclude that
the steady state is the uniform distribution
\cite{Lambiotte2008arxiv}.

For arbitrary interevent time distributions, we cannot usually calculate $r_{j \gets i}(t)$. In addition, $r_{j \gets i}(t)$ may not be symmetric with respect to $i$ and $j$ so that the steady state may deviate from the uniform distribution. To calculate the steady state in this case,
we define $f(t;j \gets i)$ as the rate at which the walker
transits from $i$ to $j$ after time $t$ has elapsed since the walker arrived at $i$. This event happens when,
link $(ij)$ is activated at time $t$ and any other link $(ik)$, where $k\neq j$, has not been activated by $t$.
Because all $\tau_{(ik)}$'s, with the case $k=j$ included, are reinitialized at the arrival of the walker at node $i$, we obtain 
\begin{equation}
\label{eq:f(t;j <- i)}
f(t;j \gets i) =
\psi_{(ij)}(t) \prod_{k \neq j;(ik)\in E } \int_t^{\infty} \psi_{(ik)}(t') dt'.
\end{equation}

We use Eq.~\eqref{eq:f(t;j <- i)} to derive the master equation for the active random walk. The rate at which the random walker reaches node $j$ from an adjacent node $i$ at time $t$ satisfies
\begin{equation}
q_{j \gets i}(t) = \int_0^t f(t-t';j \gets i) q_i(t') dt' + p_{j \gets i}(0) \delta(t),
\label{eq:q}
\end{equation}
where
\begin{equation}
q_i(t):=\sum_{k;(ki) \in E} q_{i \gets k}(t)
\label{eq:def q_i}
\end{equation}
is the rate at which the walker arrives at node $i$ at time $t$ from an arbitrary neighbor, $p_{j \gets i}(0)$ are initially chosen weights on the links satisfying 
\begin{equation}
\label{eq:initial}
\sum_{j;(ji) \in E} p_{i \gets j}(0) = p_i(0),
\end{equation}
and $\delta(t)$ is Dirac's delta function.
A detailed proof for Eq.~\eqref{eq:q} is found in Ref.~\cite{Hoffmann2012PhysRevE}. By substituting Eq.~\eqref{eq:q} in Eq.~\eqref{eq:master}, we obtain
\begin{equation}
\label{eq:intdiff} 
\frac{d}{dt}p_i(t) =\sum_{j;(ji) \in E} \int_0^t \left[ f(t-t';i \gets j) q_j(t') - f(t-t'; j \gets i) q_i(t') \right] dt'
\end{equation}
for any $t>0$. We define
\begin{equation}
f(t;i) := \sum_{j;(ji) \in E} f(t;j \gets i),
\end{equation}
i.e., the PDF of the time to transit from node $i$ to somewhere. In other words, $f(t;i)$ is the PDF of $\min_{j; (ij)\in E} \tau_{(ij)}$, which is the first time at which a link incident to $i$ is activated.
By integrating Eq.~\eqref{eq:intdiff} and abbreviating 
\begin{equation}
\label{eq:phidef}
\phi_i(t) := \int_t^\infty f(t';i) dt',
\end{equation}
which is the probability to remain at $i$ for a time longer than $t$, we obtain
\begin{equation}
p_i(t) = \int_0^t \phi_i(t-t') q_i(t') dt',
\label{eq:masterp}
\end{equation}
for any $t>0$. The derivation of Eq.~\eqref{eq:masterp} is shown in the Appendix.

\subsection{Steady state}\label{sub:steady state active}

\subsubsection{General case}

The steady state of the active random walk is evaluated via the Laplace transform of Eqs.~\eqref{eq:q} and \eqref{eq:masterp}, expansion of the exponential, and application of the final value theorem \cite{Hoffmann2012PhysRevE}.
Here we briefly present a slightly modified derivation of the steady state. We take the Laplace transform of Eq.~\eqref{eq:masterp} to obtain
\begin{equation}
\label{eq:masterplap}
\hat{p}_i(s) = \hat{\phi}_i(s) \hat{q}_i(s).
\end{equation}
Here, $\hat{p}_{i}(s)=\int_0^\infty p_{i}(t) e^{-st} dt$ is the Laplace transform of $p_{i}(t)$, and parallel definitions are applied to $\hat{\phi}_i(s)$ and $\hat{q}_i(s)$. Equation~\eqref{eq:phidef} implies
\begin{equation}
\label{eq:phi}
\hat{\phi}_i(0) = \int_0^\infty \int_t^\infty f(t';i) dt' dt = \int_0^\infty t f(t;i) dt = \left< \min_{\ell;(i\ell) \in E} \tau_{(i\ell)} \right>.
\end{equation}
According to the final value theorem, the steady state probability for node $i$, denoted by $p_i^*$, is given by $p_i^* = \lim_{s \to 0} s\hat{p}_i(s)$. By combining Eqs.~\eqref{eq:masterplap} and ~\eqref{eq:phi}, we obtain
\begin{equation}
\label{eq:steadystateactive}
p_i^* = \left< \min_{\ell;(i\ell) \in E} \tau_{(i\ell)} \right> q_i^*,
\end{equation}
where $q_i^*:=\lim_{t \to \infty} q_i(t)$ is the rate at which the random walker arrives at node $j$ in the steady state.

To calculate $\bm q^*:=(q_1^*, \ldots, q_N^*)^{\top}$, we define the $N \times N$ matrices 
\begin{equation}
\label{eq:mathbbF}
F_{\rm a}(t):=\left( f(t;j \gets i) \right)_{ji},
\end{equation}
and $\mathbb{F}_{\rm a}:=\hat{F}_{\rm a}(0)$, where $\hat{F}_{\rm a}(s) = \int_0^{\infty} F_{\rm a}(t) e^{-st} dt$ is the Laplace transform of matrix 
$F_{\rm a}(t)$. We transform Eq.~\eqref{eq:q} into the Laplace space to obtain
\begin{equation}
\label{eq:masterqlap}
\hat{\bm q}(s) = \hat{F}_{\rm a}(s) \hat{\bm q}(s) + \bm p(0),
\end{equation}
where $\hat{\bm q}(s)=(\hat{q}_1(s),\ldots,\hat{q}_N(s))^{\top}$ and $\bm p(0)=(p_1(0),\ldots,p_N(0))^{\top}$. We multiply both sides of Eq.~\eqref{eq:masterqlap} with $s$ and obtain
\begin{equation}
\label{eq:masterqlaps}
s\hat{\bm q}(s) = \hat{F}_{\rm a}(s) \left[s\hat{\bm q}(s)\right] + s\bm p(0).
\end{equation}
By taking the limit $s \rightarrow 0$ on both sides of Eq.~\eqref{eq:masterqlaps}, we find that vector $\bm q^*:=(q_1^*, \ldots, q_N^*)^{\top}$ is the dominant eigenvector of $\mathbb{F}_{\rm a}$, i.e.,
\begin{equation}
\label{eq:domeig}
\bm q^* = \mathbb{F}_{\rm a} \bm q^*.
\end{equation}
Suppose that the random walker just arrived at a node. $\mathbb{F}_{\rm a}$ contains the probabilities to make a transition from one node to another in one step, $\mathbb{F}_{\rm a}^2$ contains the probabilities to do so in two steps, and so on. Equation~\eqref{eq:domeig} implies that $\bm q^*$ is proportional to the steady state of the discrete-time random walk on the aggregate network with transition probabilities defined by $\mathbb{F}_{\rm a}$. It should be noted that $\bm q^*$ is unique up to the scaling factor because the aggregate network has been assumed to be connected. To obtain $\bm p^*$ in Eq.~\eqref{eq:steadystateactive}, we weight the steady state vector of the discrete-time random walk with the mean time for which the walker stays at the node.

\subsubsection{Identical distributions}

When interevent times for different links are identically distributed according to $\psi(t)$, 
Eq.~\eqref{eq:mathbbF} is reduced to
\begin{equation}
\left(\mathbb{F}_{\rm a}\right)_{ji}=
\begin{cases}
1/d_i & \textrm{if } (ij) \in E,\\
0 & \textrm{if } (ij) \notin E,
\end{cases}
\label{eq:F_a exponential}
\end{equation}
where $d_i$ is the degree of node $i$.
By combining Eqs.~\eqref{eq:domeig} and \eqref{eq:F_a exponential}, we obtain
\begin{equation}
\label{eq:q_i^* active identical}
q_i^* \propto d_i,
\end{equation}
which is consistent with the fact that the steady state of the simple random walk on an arbitrary static undirected network is proportional to the degree
\cite{Doyle1984book,Lovasz1993Boyal}.
By substituting Eq.~\eqref{eq:q_i^* active identical} in
Eq.~\eqref{eq:steadystateactive} and using $\sum_{i=1}^N p_i^*=1$,
we obtain
\begin{equation}
\label{eq:p_i^* active}
p_i^* = \dfrac{\langle \min_{\ell=1,\dots,d_i} \tau_{\ell} \rangle d_i}{\sum_{j=1}^N \langle \min_{\ell=1,\ldots,d_j} \tau_{\ell} \rangle d_j},
\end{equation}
where $\tau_\ell$'s are i.i.d. copies of the interevent time. It should be noted that
\begin{equation}
\label{eq:actmin}
\left< \min_{\ell=1,\dots,d_i} \tau_{\ell} \right> = \int_0^\infty \left[ \int_t^\infty \psi(t') dt' \right]^{d_i} dt
\end{equation}
depends solely on the degree of a node. Therefore, the steady state depends only on the node's degree.

If the interevent time is exponentially distributed, we obtain $\langle \min_{\ell=1,\dots,d_i} \tau_{\ell} \rangle \propto 1/d_i$ so that the steady state is the uniform distribution. This is consistent with our previous argument in Sec.~\ref{sub:formulate active}, where we derived this fact directly from the master equation [see Eq.~\eqref{eq:master}]. Otherwise, $\langle \min_{\ell=1,\dots,d_i} \tau_{\ell} \rangle$ is not necessarily proportional to $1/d_i$ such that the steady state may not be the uniform distribution.

\subsection{Mean recurrence time}\label{sub:mean recurrence time active}

Let~$T_{i|i}$ be the recurrence time, i.e., the time at which a random walker starting at $i$ returns to $i$ for the first time. We denote the PDF of $T_{i|i}$ by $g(t;i|i)$. Our goal in this section is to determine the mean recurrence time given by
\begin{equation}
\label{eq:defmrt}
\langle T_{i|i} \rangle := \int_0^\infty t g(t;i|i) dt.
\end{equation}
The hopping rate of the walker generally depends on the time already spent at a node. Therefore, we confine ourselves to the recurrence time since the walker has just arrived at node $i$. We will adapt the derivation of the mean recurrence time for discrete-time random walks on static undirected networks \cite{Noh2004PhysRevLett} to the case of the active random walk.

Denote by $p_{i|i}(t)$ the probability that the random walker is located at node $i$ at time $t$ given it started at node $i$. We obtain
\begin{equation}
\label{eq:p_ii self consistent}
p_{i|i}(t)= \phi_i(t) + \int_0^t g(t';i|i)  p_{i|i}(t-t') dt'.
\end{equation}
The first term on the right-hand side of Eq.~\eqref{eq:p_ii self consistent}
governs the case in which the random walker has not left $i$ until time $t$. The second term accounts for the walker that has left $i$ at least once.
By transforming Eq.~\eqref{eq:p_ii self consistent} into the Laplace space, we obtain
\begin{equation}
\label{eq:p_ii(s)}
\hat{p}_{i|i}(s)= \hat{\phi}_i(s) + \hat{g}(s;i|i) \hat{p}_{i|i}(s).
\end{equation}
Equation~\eqref{eq:p_ii(s)} implies
\begin{align}
\label{eq:fhat}
\hat{g}(s;i|i) =& \frac{\hat{p}_{i|i}(s)-\hat{\phi}_i(s)}{ \hat{p}_{i|i}(s)}\notag\\
=& \frac{p^*_i + sR_{ii}(s) -s \hat{\phi}_i(s)} {p^*_i + sR_{ii}(s)},
\end{align}
where
\begin{equation}
R_{ii}(s) := \hat{p}_{i|i}(s)-p^*_i/s.
\label{eq:def R_ii(s)}
\end{equation}

Equation~\eqref{eq:defmrt} implies that the mean recurrence time of node $i$ is the first moment of $g(t;i|i)$. Therefore, the mean recurrence time is derived by the Laplace transform of $g(t;i|i)$ as follows:
\begin{equation}
\label{eq:T_ii=-dg(0,ii)} 
\langle T_{i|i} \rangle =-\frac{d}{ds} \hat{g}(0,i|i).
\end{equation}
By substituting Eq.~\eqref{eq:fhat} in Eq.~\eqref{eq:T_ii=-dg(0,ii)}, we obtain 
\begin{equation}
\label{eq:mrt2}
\langle T_{i|i} \rangle =  \dfrac{\left[\hat{\phi}_i(s)+s\hat{\phi}_i'(s)\right]\left[ p^*_i + sR_{ii}(s) \right] + s \hat{\phi}_i(s) \left[ R_{ii}(s)+sR_{ii}'(s) \right]}{\left[ p^*_i + sR_{ii}(s) \right]^2} \Bigg|_{s=0}.
\end{equation}
Application of the final value theorem yields
\begin{equation}
\label{eq:sR_ii}
\lim_{s \to 0}  sR_{ii}(s) = \lim_{t \rightarrow \infty} p_{i|i}(t) - p_i^* = 0.
\end{equation}
By using the rule of L'Hospital, we obtain
\begin{equation}
\label{eq:L'Hospital to s^2 R_ii}
\lim_{s \to 0} s^2 R_{ii}'(s) = - \lim_{s \to 0} sR_{ii}(s) = 0.
\end{equation}
%
Furthermore, we recall that $\hat{\phi}_i(0)=\langle \min_{\ell \neq i;(i\ell) \in E} \tau_{(i\ell)} \rangle < \infty$ [see Eq.~\eqref{eq:phi}], so that the tail of $\phi_i(t)$ decreases fast enough to imply
\begin{equation}
\label{eq:shatphi}
\lim_{s \to 0} s\hat{\phi}_i'(s) = - \lim_{t \to \infty} t \phi_i(t) = 0.
\end{equation}
By substituting Eqs.~\eqref{eq:phi}, \eqref{eq:steadystateactive}, \eqref{eq:sR_ii}, \eqref{eq:L'Hospital to s^2 R_ii}, and \eqref{eq:shatphi} in Eq.~\eqref{eq:mrt2}, we obtain
\begin{equation}
\label{eq:mrtact}
\langle T_{i|i} \rangle = \frac{ \langle \min_{\ell ;(i\ell) \in E} \tau_{(i\ell)} \rangle}{p^*_i}= \frac{1}{q^*_i}.
\end{equation}
In particular, if interevent times on different links are identically distributed, the combination of 
Eqs.~\eqref{eq:p_i^* active} and \eqref{eq:mrtact}
yields
\begin{equation}
\langle T_{i|i} \rangle =  \frac{\sum_{j=1}^N \langle \min_{\ell=1,\ldots,d_j} \tau_{\ell} \rangle d_j}{d_i} \propto \frac{1}{d_i}.
\label{eq:T_{i|i} active identical}
\end{equation} 
It should be noted that for discrete-time random walks on undirected static networks the mean recurrence time is also inversely proportional to the node's degree \cite{Noh2004PhysRevLett}. 

\section{Passive random walk}\label{sec:passive}

In this section, we evaluate the steady state and mean recurrence time of the passive random walk. 
The main results when the interevent times of different links are distributed according to a common PDF are shown in Eqs.~\eqref{eq:ststidpas} and \eqref{eq:mrtpas} for the steady state and mean recurrence time, respectively.

The difference from the active random walk is that the move of a  walker does not reinitialize interevent times except on the link used for that jump. When the passive random walker arrives at a node, links incident to the node have thus already been inactive for some random time. 
Therefore, in contrast to the case of the active random walk, the intervent time and the waiting time $\tilde{\tau}_e$ ($e\in E$), i.e., the time until link $e$ is activated since the random walker arrives at a node incident to $e$, are different in general. They are equivalent only when the interevent time distribution is exponential, corresponding to the fact that the Poisson process is memoryless. Otherwise, the waiting time depends on the time at which the random walker arrives at a node. This distinction is important because the waiting time plays a direct role in diffusion on networks, whereas
the interevent time plays only an indirect role.

When the arrival of a walker on a node and the activation of an link are independent processes, it is known that the waiting time distribution $\rho_e(t)$ is given in terms of the interevent time distribution $\psi_e(t)$ by
\begin{equation}
\label{eq:waitapp}
\rho_e(t)= \frac{1}{{\langle \tau_e \rangle}} \int_t^\infty \psi_e(t') dt'
\end{equation}
and that the average waiting time depends on the variance of the interevent time, a property called the waiting time paradox or bus paradox in queuing theory \cite{Allen1990}. In general, the time at which the random walker jumps to an adjacent node depends on the previous trajectory of the walker. However, we assume that Eq.~\eqref{eq:waitapp} holds true in the following analysis.

\subsection{Probability flows}

In the following, we perform an approximation in order to analytically evaluate
the waiting time, steady state, and mean recurrence time of the passive random walk.
To this end, we neglect the trajectory of the random walker except for its last and current positions denoted by $k$ and $i$, respectively. More precisely, we retain the last activation time of the link through which the walker arrived at $i$ and suppose that all the other links were activated at random times in the past. 

The waiting time for link $(ik)$ is given by the interevent time distribution $\psi_{(ik)}(t)$. The waiting times for all the other links incident to $i$ are approximately distributed according to Eq.~\eqref{eq:waitapp}. The PDF of the time when the walker transits from node $i$ to node $j$ given that it arrived at node $i$ from node $k$ is approximated by
\begin{equation}
\label{eq:approx T}
f(t;j \gets i | i \gets k) \approx 
\begin{cases}
\rho_{(ij)}(t) \prod_{\ell \neq j,k;(i\ell) \in E} \left[ \int_{t}^{\infty} \rho_{(i\ell)}(r) dr \right] \int_{t}^{\infty} \psi_{(ik)}(r) dr & (j \neq k),\\
 \psi_{(ij)}(t) \prod_{\ell \neq j,k;(i\ell) \in E} \left[ \int_{t}^{\infty} \rho_{(i\ell)}(r) dr \right] & (j = k).
\end{cases}
\end{equation}
Equation \eqref{eq:approx T} suggests that the trajectory of the random walker
impacts the order of link activation. In particular, when the interevent time obeys a long-tailed distribution, a transition from node $i$ back to node $k$ is more likely than that from $i$ to another node. It should be noted that the probability of transition does not depend on the destination node in the case of the active random walk. It should be also noted that Eq.~\eqref{eq:approx T} is exact for exponentially distributed interevent times.

Similarly to Eq.~\eqref{eq:q}, we obtain
\begin{equation}
\label{eq:qpas}
q_{j \gets i}(t) \approx \sum_{k ;(ki) \in E} \left[ \int_0^t f(t-t';j \gets i|i \gets k) q_{i \gets k}(t') dt' \right] + p_{j \gets i}(0) \delta(t), 
\end{equation}
where the initial condition satisfies Eq.~\eqref{eq:initial}. Similarly to Eq.~\eqref{eq:phidef}, we define
\begin{equation}
\label{eq:pasphi}
\phi_{j \gets i}(t) := \sum_{k ; (jk) \in E} \int_t^\infty f(t';k \gets j|j \gets i) dt',
\end{equation}
which is the approximate probability that the walker stays at node $j$ for time longer than $t$ given that it arrived from node $i$. By substituting Eq.~\eqref{eq:qpas} in Eq.~\eqref{eq:master}, using Eq.~\eqref{eq:pasphi}, and performing a calculation similar to the derivation of Eq.~\eqref{eq:masterp} in the Appendix, we obtain
\begin{equation}
\label{eq:masterppas}
p_i(t) \approx \sum_{j; (ji) \in E} \left[ \int_0^t \phi_{i \gets j}(t-t') q_{i \gets j}(t') dt' \right]
\end{equation}
for any $t >0$.
Equations~\eqref{eq:qpas} and~\eqref{eq:masterppas} govern the dynamics of the passive random walk. 

\subsection{Steady state}

\subsubsection{General case}

To calculate the approximate steady state distribution,
we proceed similarly to the case of the active random walk.
%
%
By taking the Laplace transform of Eq.~\eqref{eq:qpas}, we obtain
\begin{equation}
\hat{q}_{j \gets i}(s) \approx \sum_{k ;(ki) \in E} \left[ \hat{f}(s;j \gets i|i \gets k) \hat{q}_{i \gets k}(s) \right] + p_{j \gets i}(0). 
\label{eq:hatq_ji(s)}
\end{equation}
In terms of the vectors, Eq.~\eqref{eq:hatq_ji(s)} is written as
\begin{equation}
\hat{\bm q}_{\rm p}(s) \approx \hat{F}_{\rm p}(s) \hat{\bm q}_{\rm p}(s) +\bm p_{\rm p}(0),
\label{eq:hatq passive vector}
\end{equation}
where $\hat{F}_{\rm p}(s)$ is the Laplace transform of
the $2|E|\times 2|E|$ matrix ($|E|$ is the number of links in the
aggregate network) given by
\begin{equation}
F_{\rm p}(t) := \left( f(t;j \gets i|\ell \gets k) \right)_{(ij),(k\ell) \in E},
\label{eq:mathbbFp}
\end{equation}
$\hat{\bm q}_{\rm p}(s)$ is the Laplace transform of the $2|E|$-dimensional column vector
$\bm{q}_{\rm p}(t):=(q_{i \gets j}(t))_{(ij)\in E}$, 
and $\bm p_{\rm p}(0)$ is the Laplace transform of
the $2|E|$-dimensional column vector $\bm{p}_{\rm p}(0):=(p_{i \gets j}(0))_{(ij)\in E}$.
In Eq.~\eqref{eq:mathbbFp}, we define $f(t;j \gets i|\ell \gets k) \equiv 0$ for $i \neq \ell$ because such a transition is impossible.

By multiplying both sides of Eq.~\eqref{eq:hatq passive vector} by $s$ and letting $s \to 0$, we obtain
\begin{equation}
\label{eq:paseig}
\bm q_{\rm p}^* \approx \mathbb{F}_{\rm p}\bm q_{\rm p}^*,
\end{equation}
where $\bm q_{\rm p}^*:=(q_{i \gets j}^*)_{(ij)\in E}
= (\lim_{t \to \infty} q_{j \gets i}(t))_{(ij)\in E}$ and $\mathbb{F}_{\rm p}=\hat{F}_{\rm p}(0)$.
To determine the steady state $p^*_i$, we transform Eq.~\eqref{eq:masterppas} to the Laplace space and multiply both sides by $s$ to obtain
\begin{equation}
\label{eq:passt}
s\hat{p}_i(s) \approx \sum_{j; (ji) \in E} \hat{\phi}_{i \gets j}(s) s\hat{q}_{i \gets j}(s).
\end{equation}
By setting $s \to 0$ in Eq.~\eqref{eq:passt}, we obtain
\begin{equation}
p^*_i \approx  \sum_{j;(ji) \in E} \hat{\phi}_{i \gets j}(0) q^*_{i \gets j}.
\end{equation}
Finally, Eq.~\eqref{eq:pasphi} implies
\begin{align}
\label{eq:minpas}
\hat{\phi}_{i \gets j}(0) =& \sum_{k;(ik) \in E} \int_0^\infty  \int_{t}^\infty f(t';k \gets i |i\gets j) dt'dt\notag\\
 =&  \int_0^\infty t \sum_{k; (ik) \in E}f(t;k \gets i|i \gets j) dt\notag\\ 
=& \left< \min_{k \neq j;(ik) \in E}\{ \tau_{(ij)}, \tilde{\tau}_{(ik)} \}\right>.
\end{align}
Therefore, $\hat{\phi}_{i \gets j}(0)$ is the (approximate) mean time for which a walker arriving at $i$ from node $j$ waits before moving to a neighbor.

In summary, the steady state is approximately given by
\begin{equation}
p_i^* \approx \sum_{j;(ij) \in E} \left< \min_{k \neq j;(ik) \in E}\{ \tau_{(ij)}, \tilde{\tau}_{ik} \}  \right> q_{i \gets j}^*,
\label{eq:steadystate pas gen}
\end{equation}
where $q_{i \gets j}^*$ is the solution of
Eq.~\eqref{eq:paseig}, and the normalization is given by
$\sum_{i=1}^N p_i^*=1$.

\subsubsection{Identical distributions}

Denote the components of $\mathbb{F}_{\rm p}$ by $\left(\mathbb{F}_{{\rm p}}\right)_{(ji),(\ell k)}$ for $(ij),(k\ell) \in E$.
When interevent times for different links are identically distributed according to $\psi(t)$, which we assume in this section,
we obtain
\begin{equation}
\sum_{(\ell k)\in E} \left(\mathbb{F}_{{\rm p}}\right)_{(\ell k),(ji)}=
\sum_{(ik)\in E} \left(\mathbb{F}_{{\rm p}}\right)_{(ik),(ji)}=
\sum_{(ik) \in E} \left(\mathbb{F}_{{\rm p}}\right)_{(ij),(ki)}=1.
\label{eq:doubly stochastic}
\end{equation}
The first equality in Eq.~\eqref{eq:doubly stochastic} follows from the fact that $\left(\mathbb{F}_{{\rm p}}\right)_{(\ell k),(ji)} > 0$, indicating that the walker moved from $j$ to $i$ and then from $\ell$ to $k$, if and only if $\ell=i$.
The second equality follows from the assumption that
$\psi_{e}(t)=\psi(t)$ for any $e \in E$.
Equation~\eqref{eq:doubly stochastic} indicates that $\mathbb{F}_{\rm p}$ is a doubly stochastic matrix. Therefore, the solution of Eq.~\eqref{eq:paseig} is given by 
$\bm q_{\rm p}^*\propto \mathbf{1}$, where $\mathbf{1}$ represents the $2|E|$-dimensional column vector whose all elements are equal to unity.
By using Eq.~\eqref{eq:def q_i}, we obtain
\begin{equation}
\label{eq:q_i identical}
q_i^* \propto d_i.
\end{equation}
This result is the same as that for the active random walk with identical interevent time distributions [see Eq.~\eqref{eq:q_i^* active identical}].

To evaluate the right-hand side of
Eq.~\eqref{eq:steadystate pas gen}, we use
\begin{align}
\label{eq:pasminexp}
\left<\min_{k \neq j;(ik) \in E}\{ \tau_{(ij)}, \tilde{\tau}_{ik} \}\right>
=& \left<\min_{k = 1 ,\ldots, d_i-1}\{ \tau, \tilde{\tau}_{k} \}\right>\notag\\
=& \int_0^\infty \left[ \int_t^\infty \psi(t') dt' \right] \left[ \int_t^{\infty} \rho(t')dt' \right]^{d_i-1} dt,
\end{align}
where $\tilde{\tau}_k$'s are i.i.d.\ copies of the waiting time distributed according to a common PDF $\rho(t)$. By substituting
\begin{equation}
\label{ddtintrho=intpsi}
\int_t^\infty \psi(t') dt' = -\langle \tau \rangle \frac{d}{dt} \int_t^{\infty} \rho(t')dt',
\end{equation}
which is derived by the combination of the definition of $\rho(t)$ [see Eq.~\eqref{eq:waitapp}] and the Leibniz integral rule, in Eq.~\eqref{eq:pasminexp}, we obtain
\begin{align}
\label{eq:<min>=..}
\left\langle \min_{k = 1 ,\ldots, d_i-1}\{ \tau, \tilde{\tau}_{k} \} \right\rangle
=& \int_0^\infty \left[ -\langle \tau \rangle \frac{d}{dt} \int_t^{\infty} \rho(t')dt' \right] \left[ \int_t^{\infty} \rho(t')dt' \right]^{d_i-1} dt\notag\\
=& \langle \tau \rangle - (d_i-1)\left\langle \min_{k = 1 ,\ldots, d_i-1}\{ \tau, \tilde{\tau}_{k} \} \right\rangle.
\end{align}
In the last equality in Eq.~\eqref{eq:<min>=..}, we used integration by parts.
Equation~\eqref{eq:<min>=..} implies
\begin{equation}
\label{eq:<min>=<tau>/d}
\left\langle \min_{k = 1 ,\ldots, d_i-1}\{ \tau, \tilde{\tau}_{k} \} \right\rangle = \frac{\langle \tau \rangle}{d_i}
\end{equation}
Therefore, the mean time that the passive random walker spends at a node before transiting to a neighboring node does not depend on $\psi(t)$ except for the dependence on $\left<\tau\right>$.
By combining Eqs.~\eqref{eq:def q_i}, \eqref{eq:steadystate pas gen}, \eqref{eq:q_i identical}, and \eqref{eq:<min>=<tau>/d} and using $\sum_{i=1}^N p_i^*=1$, we obtain
\begin{equation}
\label{eq:ststidpas}
p_i^* \approx \frac{1}{N}.
\end{equation}
Unlike for the active random walk [see Eq.~\eqref{eq:p_i^* active}], the (approximated) steady state of the passive random walk is the uniform distribution for any $\psi(t)$ and network structure. This is also consistent with the case when interevent times obey the exponential distribution, for which we derived the steady state in Sec.~\ref{sub:formulate active} directly from the master equation. 

\subsection{Mean recurrence time}

To evaluate the mean recurrence time for the passive random walk,
we denote by $T_{i \gets j|i}$ the time at which a random walker leaving node $i$ returns to $i$ through link $(ji)$ for the first time. The PDF of $T_{i \gets j|i}$ is denoted by $g(t;i \gets j|i)$. We also define $p_{i|i \gets j}(t)$ as the probability that the random walker is located at node $i$ at time $t$, given that it arrived at $i$ from $j$ at time 0. It should be noted that Bayes' rule results in
\begin{equation}
\label{eq:p_ii=}
p_{i|i}(t)=\sum_{j,(ji)\in E} p_{i|i \gets j}(t) \frac{p_{i \gets j}(0)}{p_i(0)}.
\end{equation}
The PDF of the first recurrence time satisfies
\begin{equation}
\label{eq:p_ii(t)}
p_{i|i}(t) \approx  \phi_i(t) + \sum_{j,(ji) \in E} \int_0^t g(t-t';i \gets j|i) p_{i|i \gets j}(t')dt'.
\end{equation}
Here, $\phi_i(t)$ denotes the probability that the walker resides at node $i$ for time longer than $t$ and is given by
\begin{equation}
\phi_i(t)=\sum_{j,(ji) \in E} \phi_{i \gets j}(t) \frac{p_{i \gets j}(0)}{p_i(0)}
\end{equation}
owing to Bayes' rule. By converting Eq.~\eqref{eq:p_ii(t)} to the Laplace space, we obtain
\begin{equation}
\label{eq:mfptform1}
\hat{p}_{i|i}(s) \approx \hat{\phi}_i(s) + \sum_{j,(ji) \in E} \hat{g}(s;i \gets j|i) \hat{p}_{i|i \gets j}(s).
\end{equation}

To evaluate Eq.~\eqref{eq:mfptform1}, we resort to a mean-field ansatz given by
$\hat{p}_{i|i \gets j}(s) \approx \hat{p}_{i|i}(s)$, where $(ji)\in E$.
In other words,
we neglect from which node the walker returns to $i$.
This approximation may be accurate if
the interevent times are identically distributed for different links, which 
we assume hereafter.

Under this approximation, Eq.~\eqref{eq:mfptform1} is reduced to
\begin{align}
\label{eq:mrt mean field}
\hat{p}_{i|i}(s) \approx& \hat{\phi}_i(s) + \sum_{j, (ji)\in E} \hat{g}(s;i \gets j|i) \hat{p}_{i|i}(s)\notag\\
=& \hat{\phi}_i(s) + \hat{g}(s;i|i) \hat{p}_{i|i}(s).
\end{align}
By following the same steps as in Eqs.~\eqref{eq:fhat}--\eqref{eq:mrtact}, we obtain
\begin{equation}
\label{eq:mrtpas2}
\langle T_{i|i} \rangle \approx \frac{ \left\langle  \min_{j = 1 ,\ldots, d_i-1}\{ \tau, \tilde{\tau}_j \} \right\rangle }{p_i^*}.
\end{equation}
By substituting Eqs. \eqref{eq:<min>=<tau>/d} and \eqref{eq:ststidpas} in Eq.~\eqref{eq:mrtpas2}, we obtain
\begin{equation}
\label{eq:mrtpas}
\langle T_{i|i} \rangle \approx \frac{N \langle \tau \rangle }{d_i}.
\end{equation}
It should be noted that the mean recurrence time is also inversely proportional to the degree for the active random walk [see Eq.~\eqref{eq:mrtact}].
It should be also noted that $\left<T_{i|i}\right>$ is independent of 
$\psi(t)$ for the passive random walk except for the factor $\left \langle\tau \right \rangle$, but not for the active random walk.

\section{Examples}

To illustrate the theoretical results derived in Secs.~\ref{sec:active} and \ref{sec:passive}, we analyze examples in this section.
We assume that the interevent times are identically distributed for all links according to $\psi(t)$, which is either a power-law or Weibull distribution.

\subsection{Power-law distributed interevent times}\label{sub:power}

\begin{figure}
\centering
\includegraphics[width=12cm]{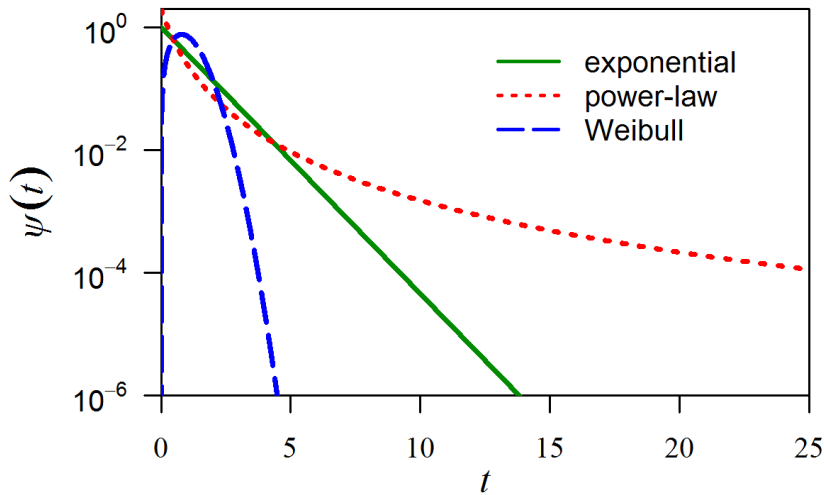}
\caption{(Color online) Three distributions of interevent times, i.e., the exponential distribution given by $\psi(t)=e^{-t}$ (solid line), the power-law distribution given by Eq.~\eqref{eq:power law psi} with $\alpha=3$ (dotted line), and the Weibull distribution given by Eq.~\eqref{eq:Weibull} with $m=2$ and $\lambda=\sqrt{\pi}/2$ (dashed line).}
\label{fig:distr}
\end{figure}

Consider the case in which all interevent times follow a power-law distribution given by 
\begin{equation}
\label{eq:power law psi}
\psi(t)=(\alpha-1)(1+t)^{- \alpha},
\end{equation}
where $\alpha > 2$, corresponding to the assumption $\left<\tau\right> < \infty$. We plot the PDFs of the power-law by the dotted line in
Fig.~\ref{fig:distr}.
In fact, many real data show $\alpha\approx 1$ or 1.5 \cite{Eckmann2004PNAS,Barabasi2005Nature,VazquezA2006PRE_burst}, which apparently contradicts our choice $\alpha>2$. Here, for simplicity we assume $\alpha>2$ to investigate the effect of long-tailed distributions on the random walk on temporal networks.

For the active random walk, the PDF of the transition time is calculated from
the substitution of Eq.~\eqref{eq:power law psi} in Eq.~\eqref{eq:f(t;j <- i)} as follows:
\begin{equation}
f(t;j \gets i)= (\alpha-1)(1+t)^{- \alpha d_i+d_i-1}. 
\end{equation}
The probability to make a transition from node $i$ to an adjacent node $j$ [see Eq.~\eqref{eq:mathbbF}] is given by
\begin{equation}
\left(\mathbb{F}_{\rm a}\right)_{ji}= \int_0^\infty f(t;j \gets i) dt 
=\frac{1}{d_i}.
\label{eq:(F_a)_ji power law}
\end{equation}
The mean time for which the random walker stays at node $i$ before moving to a
neighbor [see Eq.~\eqref{eq:phi}] is given by
\begin{equation}
\label{eq:powmean}
\left\langle \min_{\ell; (i\ell)\in E} \tau_{(i\ell)} \right\rangle = 
\frac{1}{\alpha d_i - d_i - 1}.
\end{equation}

For the passive random walk, we substitute Eq.~\eqref{eq:power law psi} in Eq.~\eqref{eq:waitapp} to obtain the PDF of the waiting time as follows: 
\begin{equation}
\label{eq:rhopow}
\rho(t)=(\alpha-2)(1+t)^{-\alpha+1}.
\end{equation}
It should be noted that the mean waiting time diverges if $2<\alpha\le 3$. However, the following results are valid for all $\alpha>2$ for the following reason. The link walked through in the last jump has a finite mean waiting time because of resetting. This fact guarantees that the mean time to the next jump is always finite. Therefore, the steady state and the mean recurrence time safely exist.
By substituting Eq~\eqref{eq:rhopow} in Eq.~\eqref{eq:approx T}, we obtain
\begin{equation}
f(t;j \gets i|i \gets k)=
\begin{cases}
(\alpha-2)(1+t)^{-\alpha d_i+2d_i-2} & \textrm{if } j \neq k,\\ 
(\alpha-1)(1+t)^{-\alpha d_i+2d_i-2} & \textrm{if } j = k. 
\end{cases}
\end{equation}
The approximate probability to make a transition from node $i$ to a neighbor $j$ conditioned that the random walker reached node $i$ through link $(ik)$ [see Eq.~\eqref{eq:mathbbFp}] is given by
\begin{equation}
\label{eq:proppow}
\left(\mathbb{F}_{{\rm p}}\right)_{(ij),(ki)}= \int_0^\infty f(t;j \gets i|i \gets k) dt =
\begin{cases}
\dfrac{\alpha-2}{\alpha d_i-2d_i+1} & \textrm{if } j \neq k,\\
\dfrac{\alpha-1}{\alpha d_i-2d_i+1} & \textrm{if } j = k.
\end{cases}
\end{equation}

To illustrate the difference between the active and passive random walks,
we consider the network composed of three nodes shown in Figs.~\ref{fig:schematic passive} and \ref{fig:schematic active}.

For the active random walk, Eq.~\eqref{eq:(F_a)_ji power law} leads to
\begin{align}
\left(\mathbb{F}_{{\rm a}}\right)_{(12),(21)} =&\left(\mathbb{F}_{{\rm a}}\right)_{(32)(23)}=1,\label{eq:transprobpowact1}\\
\left(\mathbb{F}_{{\rm a}}\right)_{(21),(12)} =&\left(\mathbb{F}_{{\rm a}}\right)_{(21),(32)}=\left(\mathbb{F}_{{\rm a}}\right)_{(23),(32)}=\left(\mathbb{F}_{{\rm a}}\right)_{(23),(12)}=\frac{1}{2}.
\label{eq:transprobpowact2}
\end{align}
These transition probabilities are the same as those in the case of
identically distributed exponential interevent times.
In contrast, for the passive random walk,
Eq.~\eqref{eq:proppow} leads to
\begin{align}
\left(\mathbb{F}_{{\rm p}}\right)_{(12),(21)} =& \left(\mathbb{F}_{{\rm p}}\right)_{(23),(32)}=1,\\
\left(\mathbb{F}_{{\rm p}}\right)_{(21),(12)} =& \left(\mathbb{F}_{{\rm p}}\right)_{(23),(32)}
= 1 - \left(\mathbb{F}_{{\rm p}}\right)_{(23),(12)}
= 1 - \left(\mathbb{F}_{{\rm p}}\right)_{(21),(32)}
\approx \frac{\alpha -1}{2\alpha-3}.
\end{align}
Because $2<\alpha<\infty$, we obtain
$0.5 < \left(\mathbb{F}_{{\rm p}}\right)_{(21),(12)}=\left(\mathbb{F}_{{\rm p}}\right)_{(23),(32)} < 1$. Therefore, the passive random walker tends to travel on the link that the walker has used the last time. In particular,
when $\alpha=2+\eps$, with $0<\eps\ll 1$, we obtain $\left(\mathbb{F}_{{\rm p}}\right)_{(21),(12)}=\left(\mathbb{F}_{{\rm p}}\right)_{(23),(32)} \approx 1 - \eps \textrm{ and } \left(\mathbb{F}_{{\rm p}}\right)_{(23),(12)}=\left(\mathbb{F}_{{\rm p}}\right)_{(21),(32)} \approx \eps$.
Therefore, a random walker starting at node 1, for example, will be 
trapped between nodes 1 and 2 for a long time before
transiting to node 3.


\begin{figure}
\centering
\includegraphics[width=10cm]{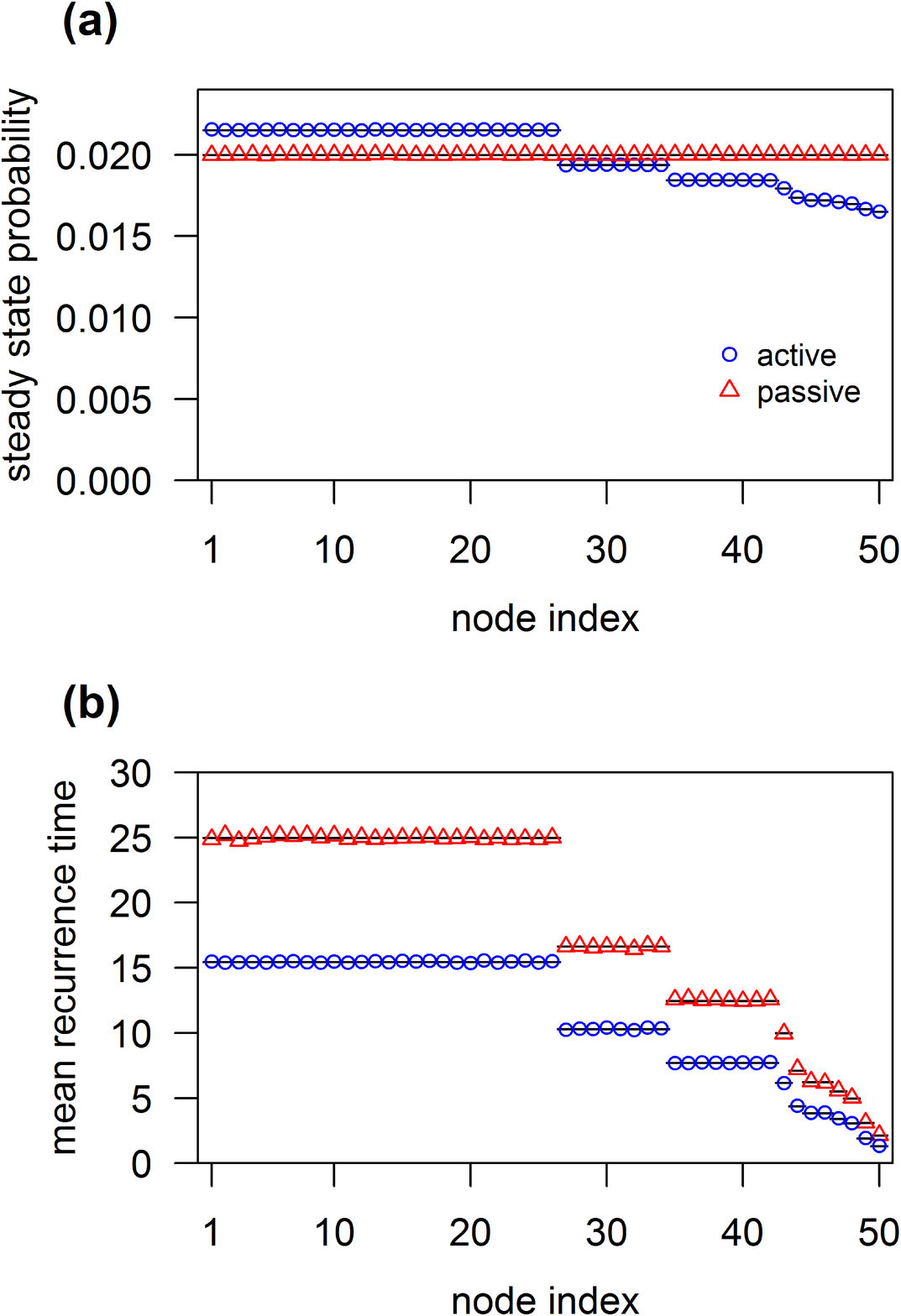}
\caption{(Color online) Numerical results for the active and passive random walk on a scale-free network with $N=50$ nodes generated by the Barab\'asi-Albert model \cite{Barabasi1999Science}. The interevent time is assumed to obey the power-law distribution with exponent $\alpha =3$ for all links. (a) Steady state of the active
(circles) and passive (triangles) random walks. (b) Mean time to return to an initial node for the active and passive random walks.
The results for the case in which the interevent time obeys 
the exponential distribution are omitted because they are indistinguishable from the results for the passive random walk (triangles). The lines represent the theoretical estimates. The nodes are sorted in ascending order of the degree.}
\label{fig:powid}
\end{figure}

For the active random walk, the mean time to stay at node $i$, i.e., Eq.~\eqref{eq:powmean}, is reduced to
\begin{equation}
\label{eq:actminid}
\left\langle \min_{\ell=1,\ldots,d_i} \tau_\ell  \right\rangle = \frac{1}{(\alpha-1)d_i - 1},
\end{equation}
where $\tau_\ell$'s are i.i.d. copies of the interevent time. By substituting Eq.~\eqref{eq:actminid} in Eq.~\eqref{eq:p_i^* active}, we obtain
\begin{equation}
\label{eq:ststactpow}
p_{{\rm a},i}^* \propto \frac{1}{\alpha-1-\frac{1}{d_i}},
\end{equation}
where subscript ``$\rm a$'' here and in the following corresponds to
the active random walk.

Equation~\eqref{eq:ststactpow} implies that the steady state is not uniform, which is in contrast to the case of exponentially distributed interevent times.
In particular, $p_{{\rm a},i}^*$ is large for nodes with small degrees, which is opposite to the case of the discrete-time simple random walk in undirected networks for which the steady state is proportional to the node's degree
\cite{Doyle1984book,Lovasz1993Boyal}.
In contrast, for the passive random walk, the steady state is the uniform distribution [see Eq.~\eqref{eq:ststidpas}].

To test our theory, we carried out numerical simulations. We set $\alpha=3$ and 
used the Barab\'asi-Albert scale-free network \cite{Barabasi1999Science} with $N=50$ nodes. The two parameters in the Barab\'asi-Albert model were set to 
$m_0=m=2$. We used the same single realization of the aggregate network for both active and passive random walks. We calculated the steady state as the average time for which
the walker spent on each node between $t=10^3$ and $t=10^8$. The choice $t=10^3$ is to exclude the transient. We initially placed the walker at one of the two nodes initially created in the
Barab\'asi-Albert algorithm, each with probability $1/2$.

The passive random walk was simulated as follows. For each link, we stored the sum of realized interevent times. As soon as the random walker arrived at a node, a new interevent time was drawn for the link just used by the walker. For each of the remaining links incident to the node, interevent times were drawn until the sum of realized interevent times exceeded the time of arrival of the random walker. Then, we selected the link $e$ with the smallest sum of realized interevent times. Finally, the random walker jumped to a new node through $e$. This procedure was repeated until the final time was reached.

The numerically obtained steady state probability of each node
is shown in
Fig.~\ref{fig:powid}(a) 
for the active (circles) and passive (triangles) random walks. 
The nodes on the horizontal axis are shown in the ascending order of the degree.
The numerical results are accurately predicted by the theory
(lines). In particular, the approximations made for analyzing the passive random walk do not cause a notable discrepancy between the numerical and theoretical results.

Next, we examine the mean recurrence time. For the active random walk, we substitute Eqs.~\eqref{eq:p_i^* active} and \eqref{eq:powmean} in Eq.~\eqref{eq:mrtact} to obtain
\begin{equation}
\langle T_{{\rm a},i|i} \rangle =  \frac{1}{d_i} \sum_{j=1}^{N} \left(\alpha-1 - \frac{1}{d_j} \right)^{-1}.
\label{eq:T_aii power law}
\end{equation}
For the passive random walk, by substituting $\left<\tau\right> = (\alpha-2)^{-1}$ in Eq.~\eqref{eq:mrtpas}, we obtain
\begin{equation}
\label{eq:T_pii power law}
\langle T_{{\rm p},i|i} \rangle \approx \frac{N}{(\alpha-2) d_i},
\end{equation}
where subscript ``$\rm p$'' corresponds to
the passive random walk.

In the numerical simulations of the passive random walk, we assume that the walker arrived at the starting node $i$ from each neighbor of $i$ with the same probability at $t=0$. To mimic the steady state of the stochastic temporal network, we assumed that the initial interevent time obeys the waiting time distribution [see Eq.~\eqref{eq:waitapp}] for all links except for link $(ji)$ from which the random walker arrived at $t=0$. The initial interevent time of link $(ji)$ is drawn from
$\psi(t)$.  For each starting node $i$, we
averaged the recurrence time over $10^5$ realizations of the random walk.

Numerically obtained mean recurrence times on the same scale-free network as that used in Fig.~\ref{fig:powid}(a) is shown in Fig.~\ref{fig:powid}(b).
The theory (lines) accurately matches the numerical results (symbols). 
We also confirmed that the numerical results when $\psi(t)$ is the exponential distribution with the same mean (i.e., $\left<\tau\right>=1$;
shown in Fig.~\ref{fig:distr} by the solid line) completely 
overlap with those for the passive random walk when $\psi(t)$ is the power-law distribution [hence not shown in Fig.~\ref{fig:powid}(b)].
Figure~\ref{fig:powid}(b) also indicates that,
for each node, the active random walk realizes a smaller mean recurrence time than the passive random walk does.

\subsection{Weibull distributed interevent times}

\begin{figure}
\centering
\includegraphics[width=10cm]{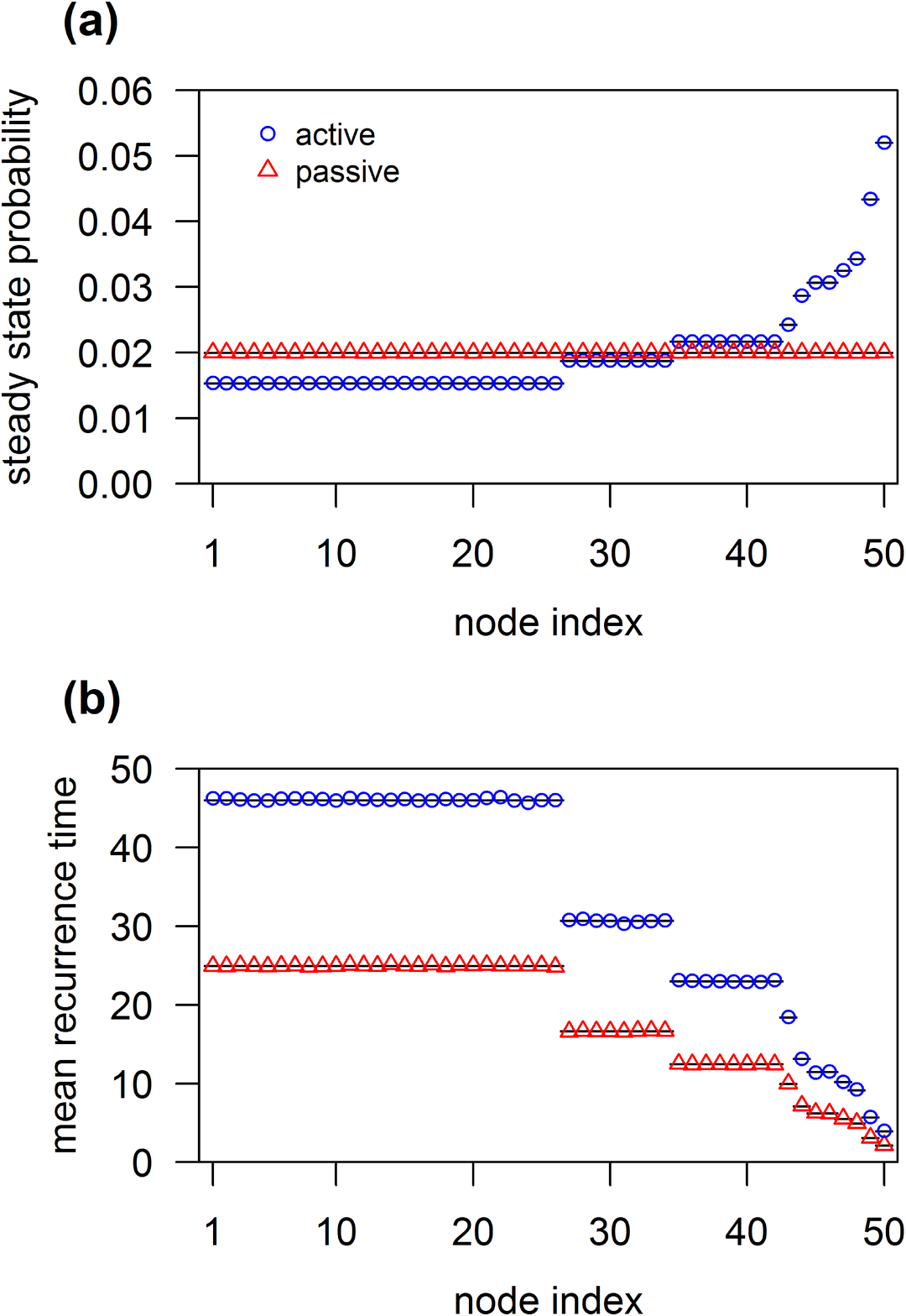}
\caption{(Color online) Numerical results for the active (circles) and passive (triangles) random walk when the interevent time obeys the Weibull distribution with $m=2$ and $\lambda=\sqrt{\pi}/2$. We used
the same realization of the scale-free network as in Fig.~\ref{fig:powid}.
(a) Steady state. (b) Mean recurrence time. The lines represent the
theory. The nodes are sorted in ascending order of the degree.}
\label{fig:weibid}
\end{figure}

For the power-law distribution of interevent times,
the steady state probability of a node decreases with the degree for the active random walk, and  the mean recurrence time at each node is larger for the passive than active random walk (Sec.~\ref{sub:power}). However, these results are not universal.

To show this, we consider the case in which the
interevent time obeys the Weibull distribution given by
\begin{equation}
\psi(t)=m\lambda^m t^{m-1} e^{-(\lambda t)^m},
\label{eq:Weibull}
\end{equation}
where $m~(0<m<\infty)$ and $\lambda (>0)$ are parameters.
The Weibull distribution with $m=2$ and $\lambda=\sqrt{\pi}/2$, which yields $\left<\tau\right>=1$, is shown by the dashed line in Fig.~\ref{fig:distr}.
It should be noted that the tail of the distribution is shorter than that of the exponential distribution with the same mean (solid line).

We start by illustrating the dynamics of the passive random walk on the 
three-node network. Refer to Fig.~\ref{fig:schematic passive} for a schematic. By combining $\rho(t)=e^{-(\lambda t)^m}$ with Eq.~\eqref{eq:approx T}, we obtain
\begin{equation}
\label{eq:F=1-2^-1/m}
\left(\mathbb{F}_{{\rm p}}\right)_{(21),(12)} = 1 - \int_0^\infty e^{-2(\lambda t)^m} dt = 1-2^{-\frac{1}{m}}
\end{equation}
independent of the $\lambda$ value.
For $m=1$, the interevent times are exponentially distributed such that $\left(\mathbb{F}_{{\rm p}}\right)_{(21),(12)}=\left(\mathbb{F}_{{\rm p}}\right)_{(23),(32)}= 1/2$. For $0<m<1$, we obtain $1/2 < \left(\mathbb{F}_{{\rm p}}\right)_{(21),(12)}=\left(\mathbb{F}_{{\rm p}}\right)_{(23),(32)} < 1$ such that the random walker tends to alternate between two nodes, which is similar to the dynamics when $\psi(t)$ is the power-law distribution (Sec.~\ref{sub:power}). For $m>1$, we obtain
$0 < \left(\mathbb{F}_{{\rm p}}\right)_{(21),(12)}=\left(\mathbb{F}_{{\rm p}}\right)_{(23),(32)} < 1/2$ such that
the random walker tends to avoid traveling on the same link in consecutive transitions. For the rest of this section, we focus on the case $m=2$. 

The numerical results for the Weibull distribution with $m=2$,
$\lambda=\sqrt{\pi}/2$,
and the same scale-free network as that used in the previous section are shown in Fig.~\ref{fig:weibid}. The steady state probability for the active random walk increases with the degree [circles in Fig.~\ref{fig:weibid}(a)].
In fact, a direct calculation of Eq.~\eqref{eq:p_i^* active} for the Weibull distribution yields $p_i^*\propto \sqrt{d_i}$, as shown by the lines overlapping with the circles in Fig.~\ref{fig:weibid}(a). This result is in contrast to the case of the power-law distribution of interevent times, for which $p_i^*$ decreases with $d_i$ [see Eq.~\eqref{eq:ststactpow}]. For the passive random walk, the steady state obeys the uniform distribution (triangles), which is consistent with the theory.

The mean recurrence time under the Weibull distribution of interevent times is shown in Fig.~\ref{fig:weibid}(b). The results for the passive random walk (triangles) are indistinguishable from those for the power-law and exponential distributions, which is consistent with the theory.
We also find that, for any node, the mean recurrence time is larger for the active than passive random walk. This result is opposite to that for the power-law distribution of interevent times.


\section{Conclusions}

We studied two models of random walks on stochastic temporal networks. Our main findings are summarized as follows. First, the steady state for the passive random walk with identically distributed interevent times on links
is uniform for any network and distribution of interevent times.
Second, for the active random walk, the steady state probability decreases and increases with the degree for the power-law and Weibull distribution of interevent times, respectively. Third, the mean recurrence time for both types of walks
is inversely proportional to the node's degree. Fourth, the mean recurrence time for the passive random walk does not depend on the distribution of interevent times. Fifth, the active random walk produces smaller mean recurrence times for each node than the passive walk does when the interevent time obeys the power-law distribution. In contrast, the mean recurrence times are larger for the active random walk than the passive random walk when the interevent time obeys the Weibull distribution.

The present result that the mean recurrence time is inversely proportional to the node's degree is consistent with that in Ref.~\cite{Starnini2012PRE}. In particular, both studies conclude that the distribution of interevent times does not affect the mean recurrence time (squares and diamonds in Fig.~7 in Ref.~\cite{Starnini2012PRE}). We reached this conclusion by explicit derivation of the mean recurrence time. In contrast, we consider that
the strength of Ref.~\cite{Starnini2012PRE} in this respect lies in numerically showing the universality of this result across different data sets. It should be noted that a discrete-time simple random walk on a different temporal network model yields different results; the mean recurrence time decreases but is not inversely proportional to the degree \cite{Perra2012PhysRevLett}.

The passive random model induces a correlated random walk. Interesting connections of the present study may be made to seminal work on correlated random walk on lattices \cite{Goldstein1951QuartJMechApplMath,Gillis1955ProcCambPhilosSoc,Klein1952} and to recent work modeling empirical pathways on networks by second-order Markov processes \cite{REL13,Scholtes2013,Lambiotte2014}. Pursuing connection to anomalous diffusion on lattices \cite{Balescu1997} may be also interesting.
It may be also interesting to explore the case in which the mean interevent time diverges, which would make the walker to linger at a node for extremely long time.

\section*{Acknowledgments}
L.S. acknowledges the support provided through ERATO, JST and DAAD. R.L. acknowledges financial support from FNRS, from the ARC ``Mining  
and Optimization of Big Data Models'', and from the EU project  
Optimizr. This paper presents research results of the Belgian Network  
DYSCO funded by the Interuniversity Attraction Poles Programme. N.M. acknowledges the support provided through CREST, JST. This work was partly supported by a Bilateral Joint Research Project between JSPS, Japan, and F.R.S.-FNRS, Belgium.

\appendix

\section{Derivation of Eq.~\eqref{eq:masterp}}\label{sec:derivation}

By substituting Eq.~\eqref{eq:q} in Eq.~\eqref{eq:master}, we obtain
\begin{equation}
\label{eq:masterappend}
\frac{d}{dt}p_i(t) = q_i(t) - \sum_{j;(ij) \in E} \int_0^t f(t-t';j \gets i)q_i(t') dt'
\end{equation}
for any $t >0$ and $1 \le i \le N$.
By integrating Eq.~\eqref{eq:masterappend}, we obtain
\begin{equation}
p_i(t)-p_i(0)
=\int_0^t \left[ q_i(t') - \sum_{j;(ij) \in E} \int_0^{t'} f(t'-t'';j \gets i)q_i(t'') dt'' \right] dt'.
\label{app:int}
\end{equation}
By applying the Leibniz integral rule, i.e.,
\begin{equation}
\frac{d}{dt} \int_{\alpha(t)}^{\beta(t)} f(t;t') dt' = f(t;\beta(t)) \frac{d}{dt} \beta(t) - f(t;\alpha(t)) \frac{d}{dt} \alpha(t) + \int_{\alpha(t)}^{\beta(t)} \frac{d}{dt} f(t;t') dt',
\end{equation}
%
%
by setting $\alpha(t)=0$, $\beta(t)=t$, and $f(t;t')=\phi_i(t-t')q_i(t')$,
we obtain
\begin{equation}
\label{app:form1}
\frac{d}{dt} \int_{0}^{t} \phi_i(t-t')q_i(t') dt' 
= \phi_i(0)q_i(t) + \int_{0}^{t} \frac{d}{dt} \phi_i(t-t')q_i(t') dt'.
\end{equation}
To evaluate the second term on the right-hand side of Eq.~\eqref{app:form1}, we use the definition of $\phi_i(t)$ [see Eq.~\eqref{eq:phidef}] to obtain
\begin{equation}
\label{app:form2}
\frac{d}{dt} \phi_i(t-t') = -\frac{d}{dt} \int_{t-t'}^\infty \sum_{j;(ij) \in E} f(t'';j \gets i) dt'' = - \sum_{j;(ij) \in E} f(t-t';j \gets i).
\end{equation}
By using $\phi_i(0)=1$ and
substituting Eq.~\eqref{app:form2} in Eq.~\eqref{app:form1}, we obtain
\begin{equation}
\label{app:form3}
\frac{d}{dt} \int_{0}^{t} \phi_i(t-t')q_i(t') dt' 
= q_i(t) - \sum_{j;(ij) \in E} \int_{0}^{t}  f(t-t';j \gets i)q_i(t') dt'.
\end{equation}
By substituting Eq.~\eqref{app:form3} in Eq.~\eqref{app:int}, 
we obtain
\begin{equation}
\label{app:form4}
p_i(t)-p_i(0) = \int_0^t \frac{d}{dt'} \int_{0}^{t'} \phi_i(t'-t'')q_i(t'') dt'' dt'.
\end{equation}
To evaluate the right-hand side of Eq.~\eqref{app:form4}, we have to evaluate
\begin{equation}
\label{app:initialcondition}
\lim_{t \to 0} \int_{0}^{t} \phi_i(t-t')q_i(t') dt'.
\end{equation}
This task needs carefulness because
$q_i(t)$ behaves like $p_i(0) \delta(t)$ around $t=0$.
By using the initial value theorem, we obtain 
\begin{equation}
\label{app:limint}
\lim_{t \to 0} \int_{0}^{t} \phi_i(t-t')q_i(t') dt' = \lim_{s \rightarrow \infty} s\hat{\phi}_i(s)\hat{q}_i(s).
\end{equation}
Because $\lim_{s \rightarrow \infty} s\hat{\phi}_i(s)=\phi_i(0)=1$, we obtain
\begin{equation}
\label{app:limq}
\lim_{s \rightarrow \infty} s\hat{\phi}_i(s)\hat{q}_i(s) = \lim_{s \rightarrow \infty} \hat{q}_i(s)=\lim_{s \rightarrow \infty} \sum_{j;(ji)\in E} \hat{q}_{i \gets j}(s),
\end{equation}
where the last equality follows from Eq.~\eqref{eq:def q_i}.
The Laplace transform of Eq.~\eqref{eq:q} yields
\begin{equation}
\label{app:limsum}
\lim_{s \rightarrow \infty} \sum_{j;(ji)\in E} \hat{q}_{i \gets j}(s) =  \lim_{s \rightarrow \infty}\sum_{j;(ji)\in E} \hat{f}(s;i \gets j)\hat{q}_j(s)+p_i(0).,
\end{equation}
Using the dominated convergence theorem, we obtain
\begin{equation}
\lim_{s \rightarrow \infty}\hat{f}(s;i \gets j)= \lim_{s \to \infty} \int_0^\infty f(t;i \gets j)e^{-ts} dt = \int_0^\infty \lim_{s \to \infty} f(t;i \gets j)e^{-ts} dt=0.
\label{app:limsum2}
\end{equation}
The use of the dominated convergence theorem is justified because 
\begin{equation}
f(t;i \gets j)e^{-ts} < f(t;i \gets j)
\end{equation}
and
\begin{equation}
\int_0^\infty f(t;i \gets j)dt < \infty.
\end{equation}
By combining Eqs.~\eqref{app:limint}, \eqref{app:limq}, \eqref{app:limsum}, and \eqref{app:limsum2}, we obtain
\begin{equation}
\lim_{t \to 0} \int_{0}^{t} \phi_i(t-t')q_i(t') dt' = p_i(0).
\end{equation}
Therefore, by evaluating the right-hand side of Eq.~\eqref{app:form4}, we obtain
\begin{equation}
p_i(t)-p_i(0) = \int_0^t  \phi_i(t-t')q_i(t') dt - p_i(0),
\end{equation}
which is equivalent to Eq.~\eqref{eq:masterp}.

\end{document}